%% file: 00paper.tex
\documentclass[sigconf]{acmart}

\usepackage{booktabs}

\usepackage[normalem]{ulem}  % for \sout => should be removed later
\usepackage{bbold}  % for \mathbb
\usepackage{subfigure}  % for \subfigure 
\usepackage{placeins}  % for \FloatBarrier
\usepackage{multirow} % for \multirow
\usepackage{moreverb} % for verbatimtab

\usepackage{tabularx} % for \tabularx

\setcopyright{rightsretained}

\copyrightyear{2017} 
\acmYear{2017} 
\setcopyright{acmcopyright}
\acmConference{SIGIR'17}{}{August 7--11, 2017, Shinjuku, Tokyo, Japan}\acmPrice{15.00}\acmDOI{http://dx.doi.org/10.1145/3077136.3080745}
\acmISBN{978-1-4503-5022-8/17/08}

\begin{document}

\title{Generating Query Suggestions to Support Task-Based Search}

% -------------
% Authors
\author{Dar\'{i}o Garigliotti}
\affiliation{University of Stavanger}
\email{dario.garigliotti@uis.no}

\author{Krisztian Balog}
\affiliation{University of Stavanger}
\email{krisztian.balog@uis.no}

% -------------

\begin{abstract}
We address the problem of generating query suggestions to support users in completing their underlying tasks (which motivated them to search in the first place). 
Given an initial query, these query suggestions should provide a coverage of possible subtasks the user might be looking for.
We propose a probabilistic modeling framework that obtains keyphrases from multiple sources and generates query suggestions from these keyphrases.  
Using the test suites of the TREC Tasks track, we evaluate and analyze each component of our model.
\end{abstract}




\maketitle

\input{sigir2017-tasks-01}

\input{sigir2017-tasks-02}
\input{sigir2017-tasks-03}

\input{sigir2017-tasks-04}

\input{sigir2017-tasks-05}
\input{sigir2017-tasks-06}

% Bibliography at the end, not splitted with figures inside
\FloatBarrier  % from {placeins} package

\bibliographystyle{ACM-Reference-Format}
\bibliography{sigir2017-tasks}

\end{document}

%% file: sigir2017-tasks-01.tex
\section{Introduction}
\label{sec:intro}

\begin{figure}[t]
    \centering
    \includegraphics[width=0.3\textwidth]{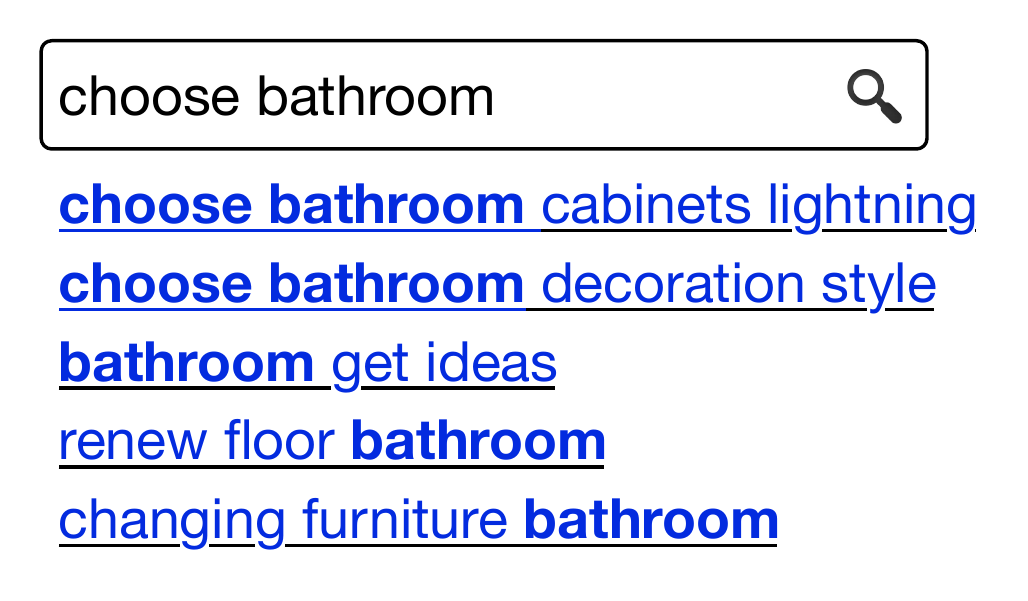}    
	\centering
	\vspace{-0.05in}
    \caption{Query suggestions to support task-based search.}
    \label{fig:ui}
\end{figure}

Search is often performed in the context of some larger underlying task~\cite{Kelly:2013:NWT}.
There is a growing stream of research aimed at making search engines more task-aware (i.e., recognizing what task the user is trying to accomplish) and customizing the search experience accordingly (see \S\ref{sec:rel}). 
In this paper, we focus our attention on one particular tool for supporting task-based search: query suggestions.
Query suggestions are an integral part of modern search engines~\cite{Ozertem:2012:LSM}.  We envisage an user interface where these suggestions are presented once the user has issued an initial query; see Figure~\ref{fig:ui}.  Note that this is different from query autocompletion, which tries to recommend various possible completions while the user is still typing the query. 
The task-aware query suggestions we propose are intended for exploring various aspects (subtasks) of the given task after inspecting the initial search results.  Selecting them would allow the user to narrow down the scope of the search.

The Tasks track at the Text REtrieval Conference (TREC) has introduced an evaluation platform for this very problem, referred to as \emph{task understanding}~\cite{Yilmaz:2015:OTT}. 
Specifically, given an initial query, the system should return a ranked list of keyphrases ``that represent the set of all tasks a user who submitted the query may be looking for''~\cite{Yilmaz:2015:OTT}.
The goal is to provide a complete coverage of subtasks for an initial query, while avoiding redundancy.
We use these keyphrases as \emph{query suggestions}.  

Our aim is to generate such suggestions in a setting where past usage data and query logs are not available or cannot be utilized.  This would be typical for systems that have a smaller user base (e.g., in the enterprise domain) or when a search engine has been newly deployed~\cite{Bhatia:2011:QSA}.
One possibility is to use query suggestion APIs, which are offered by all major web search engines.  These are indeed one main source type we consider.  Additionally, we use the initial query to search for relevant documents, using web search engines, and extract keyphrases from search snippets and from full text documents.  Finally, given the task-based focus of our work, we lend special treatment to the WikiHow site,\footnote{http://www.wikihow.com/} which is an extensive database of how-to guides.

% Overview UI figure
%

The main contribution of this paper is twofold. First, we propose a transparent architecture, using generative probabilistic modeling, for extracting keyphrases from a variety of sources and generating query suggestions from them.
Second, we provide a detailed analysis of the components of our framework using different estimation methods.   
Many systems that participated in the TREC Tasks track have relied on strategic combinations of different sources to produce query suggestions, see, e.g.,~\cite{Hagen:2015:WAT,Garigliotti:2016:UST, Hagen:2016:WAT}.  However, no systematic comparison of the different source types has been performed yet---we fill this gap.
Additional components include estimating a document's importance within a given source, extracting keyphrases from documents, and forming query suggestions from these keyphrases.
Finally, we check whether our findings are consistent across the 2015 and 2016 editions of the TREC Tasks track.

%% file: sigir2017-tasks-02.tex
\section{Related Work}
\label{sec:rel}

There is a large body of work on understanding and supporting users in carrying out complex search tasks.
Log-based studies have been one main area of focus, including the identification of tasks and segmentation of search queries into tasks~\citep{Lucchese:2013:DTS,HassanAwadallah:2014:SCS} and mining task-based search sessions in order to understand query reformulations~\citep{Jiang:2012:CEQ} or search trails~\citep{White:2010:ASR}.  
Another theme is supporting exploratory search, where users pursue an information goal to learn or discover more about a given topic.  Recent research in this area has brought the importance of support interfaces into focus~\citep{Andolina:2015:ISP,Tran:2016:FNM,Balog:2015:TCE}.
Our main interest is in query suggestions, as a distinguished support mechanism.  
Most of the related work utilizes large-scale query logs.  For example, \citet{Craswell:2007:RWC} perform a random walk on a query-click graph. \citet{Boldi:2008:QGM} model the query flow in user search sessions via chains of queries.
Scenarios in the absence of query logs have been addressed in~\citep{Kruschwitz:2013:DQS,Bhatia:2011:QSA}, where query suggestions are extracted from the document corpus.  However, their focus is on query autocompletion, representing the completed and partial terms in a query.
\citet{Kelly:2009:CQT} have shown that users prefer query suggestions, rather than term suggestions.  We undertake the task of suggesting queries to users, related to the task they are performing, as we shall explain in the next section.

%% file: sigir2017-tasks-03.tex
\section{Problem statement}
\label{sec:problem}

We adhere to the problem definition of the task understanding task of the TREC Tasks track.  
Given an initial query $q_0$, the goal is to return a ranked list of query suggestions $\langle q_1, \dots q_n \rangle$ that cover all the possible subtasks related to the task the user is trying to achieve.  
In addition to the initial query string, the entities mentioned in it are also made available (identified by their Freebase IDs).  

For example, for the query ``low wedding budget,'' subtasks include (but are not limited to) ``buy a used wedding gown,'' ``cheap wedding cake,'' and ``make your own invitations.'' These subtasks have been manually identified by the track organizers based on information extracted from the logs of a commercial search engine.
The suggested keyphrases are judged with respect to each subtask on a three point scale (non-relevant, relevant, and highly relevant).
Note that subtasks are only used in the evaluation, these are not available when generating the keyphrases.

%% file: sigir2017-tasks-04.tex
\section{Approach}
\label{sec:approach}

\begin{figure}[t]
    \centering
    \includegraphics[width=0.45\textwidth]{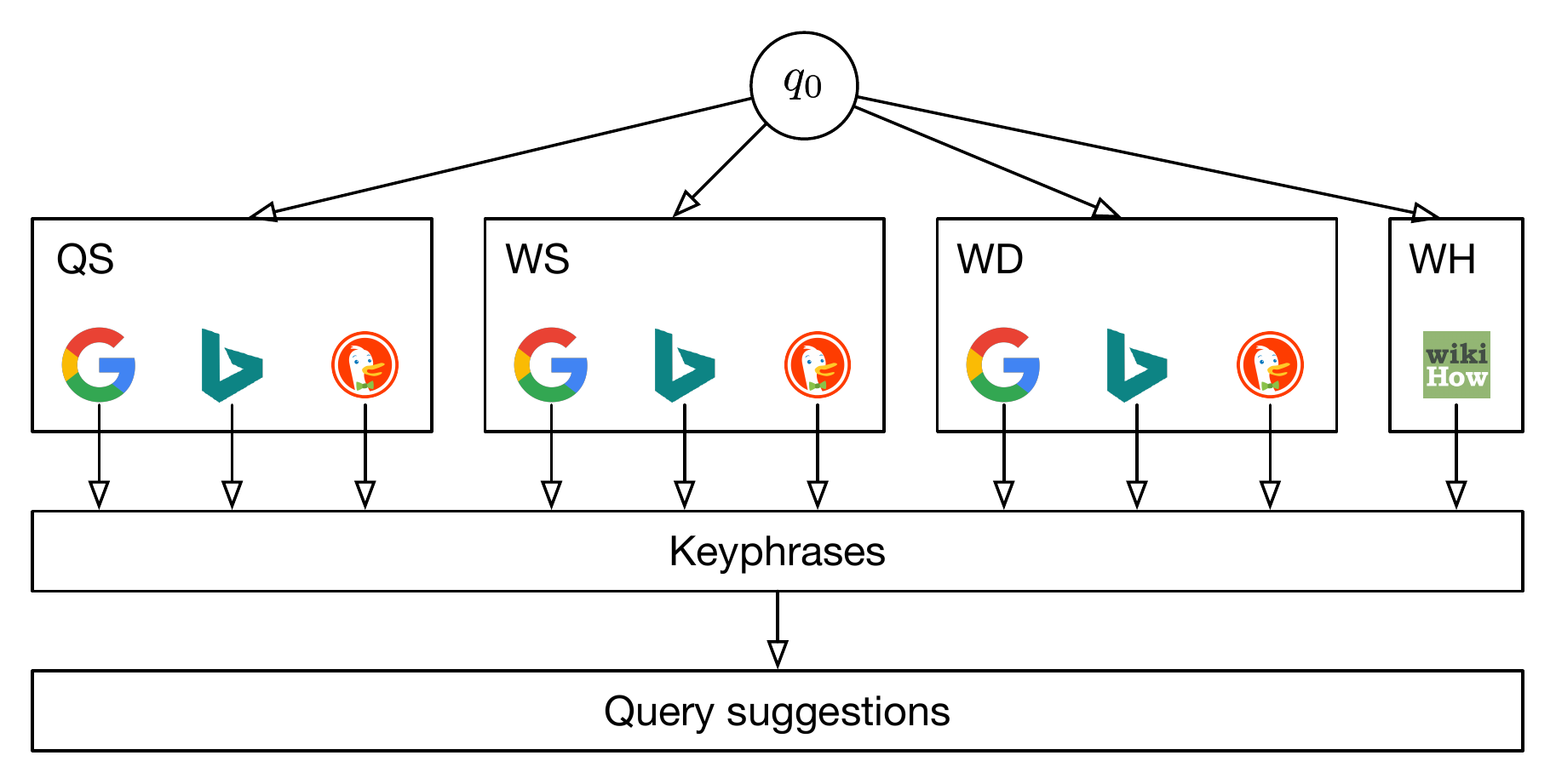}    
	\centering
    \caption{High-level overview of our approach.}
    \label{fig:architecture}
    \vspace{-0.2in}
\end{figure}

We now present our approach for generating query suggestions. 
As Figure~\ref{fig:architecture} illustrates, we obtain keyphrases from a variety of sources, and then construct a ranked list of query suggestions from these.

% High-level architecture figure
%
%

% ------------------------------------------------
\subsection{Generative Modeling Framework}

We introduce a generative probabilistic model for scoring the candidate query suggestions according to $P (q|q_0)$, i.e., the probability that a query suggestion $q$ was generated by the initial query $q_0$.
Formally:
\begin{align*}
	P(q|q_0) & = \sum_s P(q|q_0, s) P(s|q_0) \\
	& = \sum_s \big( \sum_d P(q|q_0, s, d) P(d| q_0, s) \big) P(s|q_0) \\
	& = \sum_s \big( \sum_d \Big( \sum_k P(q|q_0, s, k) P(k|s, d)  \Big) P(d| q_0, s) \big) P(s|q_0)~.
\end{align*}
This model has four components:
(i) $P(s|q_0)$ expresses the importance of a particular information source $s$ for the initial query $q_0$;
(ii) $P(d|q_0, s)$ represents the importance of a document $d$ originating from source $s$, with respect to the initial query; 
(iii) $P(k|d, s)$ is the relevance of a keyphrase $k$ extracted from a document $d$ from source $s$; and
(iv) $P(q|q_0, s, k)$ is the probability of generating query suggestion  $q$, given keyphrase $k$, source $s$, and the initial query $q_0$.
Below, we detail the estimation of each of these components.

% ------------------------------------------------
\subsection{Source Importance}
\label{sec:approach:s_imp}

We collect relevant information from four kinds of sources: query suggestions (QS), web search snippets (WS), web search documents (WD), and WikiHow (WH).  For the first three source types, we use three different web search engines (Google, Bing, and DuckDuckGo), thereby having a total of 10 individual sources.
In this work, we assume conditional independence between a source $s$ and the initial query $q_0$, i.e., set $P(s|q0) = P(s)$.

% ------------------------------------------------
\subsection{Document Importance}
\label{sec:approach:d_imp}

From each source $s$, we obtain the top-$K$ ($K$ = 10) documents for the query $q_0$.  We propose two ways of modeling the importance of a document $d$ originating from $s$: (i) uniform and (ii) inversely proportional to the rank of $d$ among the top-$K$ documents, that is:
\begin{align}
	P(d|q_0, s) & = {{K - r + 1}\over{\sum_{i=1}^{K} K - i + 1}} = {{K - r + 1}\over{K(K+1)/2}} ~~, \nonumber 
\end{align}
where $r$ is the rank position of $d$ ($r \in [1..K]$).

% ------------------------------------------------
\subsection{Keyphrase Relevance}
\label{sec:approach:k_extr}

We obtain keyphrases from each document, using an automatic keyphrase extraction algorithm.  Specifically, we use the RAKE keyword extraction system~\cite{Medelyan:2015:RMT}.
For each keyphrase $k$, extracted from document $d$, the associated confidence score is denoted by $c(k,d)$.
Upon a manual inspection of the extraction output, we introduce some data cleansing steps.  We only retain keyphrases that: (i) have an extraction confidence above an empirically set threshold of 2; (ii) are at most 5 terms long; (iii) each of the terms has a length between 4 and 15 characters, and is either a meaningful number (i.e., max. 4 digits) or a term (excluding noisy substrings and reserved keywords from mark-up languages).  Finally, we set the relevance of $k$ as $P(k|d,s) = c(k, d)/\sum_{k'} c(k', d)$.

In case $s$ is of type QS, each returned document actually corresponds to a query suggestion. Thus, we treat each of these documents $d$ as a single keyphrase $k$, for which we set $P(k|d, s) = 1$.

% ------------------------------------------------
\subsection{Query Suggestions}
\label{sec:approach:q_suggs}

As a final step, we need to generate query suggestions from the extracted keyphrases.  As a baseline option, we take each raw keyphrase $k$ as-is, i.e., with $q = k$ we set $P(q|q_0, s, k) = 1$.

Alternatively, we can form query suggestions by \emph{expanding keyphrases}. Here, $k$ is combined with the initial query $q_0$ using a set of expansion rules proposed in~\cite{Garigliotti:2016:UST}:
(i) adding $k$ as a suffix to $q_0$; (ii) adding $k$ as a suffix to an entity mentioned in $q_0$; and (iii) using $k$ as-is.
Rules (i) and (ii) further involve a custom string concatenation operator; we refer to~\cite{Garigliotti:2016:UST} for details.
Each query suggestion $q$, that is generated from keyword $k$, has an associated confidence score $c(q, q_0, s, k)$. We then set $P(q|q_0, s, k) = c(q, q_0, s, k) / \sum_{q'} c(q', q_0, s, k)$.
By conditioning the suggestion probability on $s$, it is possible to apply a different approach for each source.  Like in the previous subsection, we treat sources of type QS distinctly, by simply taking $q = k$ and setting $P(q|q_0, s, k) = 1$.  

We note that it is possible that multiple query suggestions have the same final probability $P(q|q_0)$. We resolve ties using a deterministic algorithm, which orders query suggestions by length (favoring short queries) and then sorts them alphabetically.

%% file: sigir2017-tasks-05.tex
\section{Results}
\label{sec:results}

In this section we present our experimental setup and results.

\begin{table}[t]
  \centering
  \caption{Comparison of query suggestion generators across the different types of sources. Statistical significance is tested against the corresponding line in the top block.}
  \begin{tabular}{ ll llll }
	\toprule
    $P(q|q_0, s, k)$ & $S$ & \multicolumn{2}{c}{2015} & \multicolumn{2}{c}{2016} \\
	& & ERR-IA & $\alpha$-NDCG & ERR-IA & $\alpha$-NDCG \\
    \midrule
	Using raw & QS & 0.0755 & 0.1186 & \textbf{0.4114} & \textbf{0.5289} \\
    keyphrases & WS & \textbf{0.2011} & \textbf{0.2426} & 0.3492 & 0.4038 \\
    & WD & 0.1716 & 0.2154 & 0.2339 & 0.2886 \\
    & WH & 0.0744 & 0.1044 & 0.1377 & 0.1723 \\
    \midrule
    Using & QS & 0.0751 & 0.1182 & \textbf{0.4046} & \textbf{0.5233} \\
    expanded & WS & \textbf{0.1901} & \textbf{0.2274} & 0.2927{$^\dag$} & 0.3467{$^\dag$} \\
    keyphrases & WD & 0.1551 & 0.2097 & 0.1045{$^\ddag$} & 0.1667{$^\ddag$} \\
    & WH & 0.0849 & 0.1090 & 0.0789{$^\dag$} & 0.0932{$^\ddag$} \\
    \bottomrule
  \end{tabular}
  \label{table:qry_sugg}
  \vspace{-0.02in}
\end{table}
%

% ------------------------------------------------
\subsection{Experimental Setup}
\label{sec:results:setup}

We use the test suites of the TREC 2015 and 2016 Tasks track~\cite{Yilmaz:2015:OTT,Verma:2016:OTT}.  These contain 34 and 50 queries with relevance judgments, respectively.
We report on the official evaluation metrics used at the TREC Tasks track, which are ERR-IA@20 and $\alpha$-NDCG@20.  In accordance with the track's settings, we use ERR-IA@20 as our primary metric.  (For simplicity, we omit mentioning the cut-off rank of 20 in all the table headers.)
We noticed that in the ground truth the initial query itself has been judged as a highly relevant suggestion in numerous cases. We removed these cases, as they make little sense for the envisioned scenario; we note that this leads to a drop in absolute terms of performance. 
We report on statistical significance using a two-tailed paired t-test at $p<0.05$ and $p<0.001$, denoted by $^\dag$ and $^\ddag$, respectively.

In a series of experiments, we evaluate each component of our approach, in a bottom-up fashion.
For each query set, we pick the configuration that performed best on that query set, which is an idealized scenario.  Note that our focus is not on absolute performance figures, but on answering the following research questions:   

\begin{description}
	\item[RQ1] What are the most useful information sources? % for our task?
	\item[RQ2] What are effective ways of (i) estimating the importance of  documents and (ii) generating query suggestions from keyphrases?
	\item[RQ3] Are our findings consistent across the two query sets? 
\end{description}

% ------------------------------------------------
\subsection{Query Suggestion Generation}
\label{sec:results:p_q}

We start our experiments by focusing on the generation of query suggestions and compare the two methods described in \S\ref{sec:approach:q_suggs}.  The document importance is set to be uniform.  We report performance separately for each of the four source types $S$ (that is, we set $P(s)$ uniformly among sources $s \in S$ and set $P(s)=0$ for $s \not\in S$).
Table~\ref{table:qry_sugg} presents the results.  It is clear that, with a single exception (2015 WH), it is better to use the raw keyphrases, without any expansion. The differences are significant on the 2016 query set for all source types but QS.
Regarding the comparison of different source types, we find that QS $>$ WS $>$ WD $>$ WH on the 2016 query set, meanwhile for 2015, the order is WS $>$ WD $>$ QS, WH.

% ------------------------------------------------
\subsection{Document Importance}
\label{sec:results:p_d}

\begin{table}[t]
  \centering
  \caption{Comparison of document importance estimators across the different types of sources. Statistical significance is tested against the corresponding line in the top block.}
  \begin{tabular}{ l l l l l l }
	\toprule
    $P(d|q_0, s)$ & $S$ & \multicolumn{2}{c}{2015} & \multicolumn{2}{c}{2016} \\
	& & ERR-IA & $\alpha$-NDCG & ERR-IA & $\alpha$-NDCG \\
    \midrule
	Uniform & QS & 0.0755 & 0.1186 & \textbf{0.4114} & \textbf{0.5289} \\
    & WS & \textbf{0.2011} & \textbf{0.2426} & 0.3492 & 0.4038 \\
    & WD & 0.1716 & 0.2154 & 0.2339 & 0.2886 \\
    & WH & 0.0849 & 0.1090 & 0.1377 & 0.1723 \\
    \midrule
    Rank-based & QS & 0.0891{$^\dag$} & 0.1307{$^\dag$} & \textbf{0.4288} & \textbf{0.5455} \\
    decay & WS & \textbf{0.1906} & \textbf{0.2315} & 0.3386 & 0.4011 \\
    & WD & 0.1688 & 0.2119 & 0.1964 & 0.2608 \\
    & WH & 0.0935 & 0.1225 & 0.1195 & 0.1495 \\
    \bottomrule
  \end{tabular}
  \label{table:doc_imp}
\end{table}

Next, we compare the two document importance estimator methods, uniform and rank-based decay (cf. \S\ref{sec:approach:d_imp}), for each source type.  Table~\ref{table:doc_imp} reports the results.
We find that rank-based document importance is beneficial for the query suggestion (QS) source types, for both years, and for WikiHow (WH) on the 2015 topics.  However, the differences are only significant for QS 2015.  For all other source types, the uniform setting performs better.  

\begin{figure}[t]
	\centering
	\includegraphics[width=0.44\textwidth]{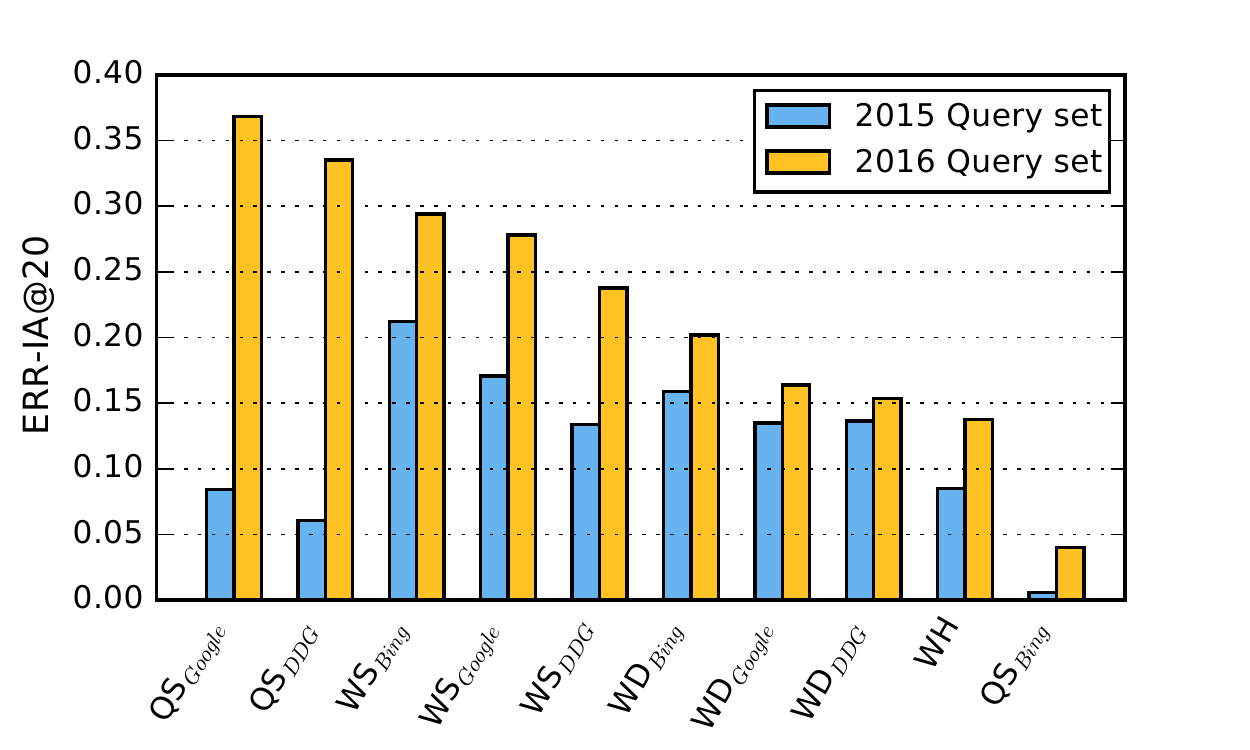}
	\vspace{-0.12in}
	\caption{Performance of individual sources, sorted by performance on the 2016 query set.}
	\label{fig:eval_individual_p_d}
\end{figure}

We also compare performance across the 10 individual sources.  
Figure~\ref{fig:eval_individual_p_d} shows the results, in terms of ERR-IA@20, using the uniform estimator.  We observe a very similar pattern using the rank-based estimator (which is not included due to space constraints).
On the 2016 query set, the individual sources follow the exact same patterns as their respective types (i.e., QS $>$ WS $>$ WD $>$ WH), with one exception.  The Bing API returned an empty set of search suggestions for many queries, hence the low performance of QS$_{Bing}$.  We can observe a similar pattern on the 2015 topics, with the exception of sources of type QS, which are the least effective here.

%

% ------------------------------------------------
\subsection{Source Importance}
\label{sec:results:p_s}

Finally, we combine query suggestions across different sources; for that, we need to set the importance of each source. 
We consider three different strategies for setting $P(s)$: 
(i) uniformly; 
(ii) proportional to the importance of the corresponding source type (QS, WS, WD, and WH) from the previous step (cf. Table~\ref{table:doc_imp});
(iii) proportional to the importance of the individual source (cf. Figure~\ref{fig:eval_individual_p_d}).
The results are presented in Table~\ref{table:src_imp}.  
Firstly, we observe that the combination of sources performs better than any individual source type on its own.
As for setting source importance, on the 2015 query set we find that (iii) delivers the best results, which is in line with our expectations.  On the 2016 query set, only minor differences are observed between the three methods, none of which are significant.

% ------------------------------------------------
\subsection{Summary of Findings}

(RQ1) Query suggestions provided by major web search engines are unequivocally the most useful information source on the 2016 queries. We presume that these search engine suggestions are already diversified, which we can directly benefit from for our task.
These are followed, in order, by keyphrases extracted from (i) web search snippets, (ii) web search results, i.e., full documents, and (iii) WikiHow articles.  On the 2015 query set, query suggestions proved much less effective; see RQ3 below.
(RQ2) With a single exception, using the raw keyphrases, as-is, performs better than expanding them by taking the original query into account. For web query suggestions it is beneficial to consider the rank order of suggestions, while for web search snippets and documents the uniform setting performs better.  For WikiHow, it varies across query sets.
(RQ3) Our main observations are consistent across the 2015 and 2016 query sets, regarding documents importance estimation and suggestions generation methods.  
It is worth noting that some of our methods were officially submitted to TREC 2016~\citep{Garigliotti:2016:UST} and were included in the assessment pools.  This is not the case for 2015, where many of our query suggestions are missing relevance assessments (and, thus, are considered irrelevant).  This might explain the low performance of QS sources on the 2015 queries.

\begin{table}[t]
  \centering
  \caption{Combination of all sources using different source importance estimators. Significance is tested against the uniform setting (line 1).}
  \begin{tabular}{ l llll }
	\toprule
    $P(s)$ & \multicolumn{2}{c}{2015} & \multicolumn{2}{c}{2016} \\
	& ERR-IA & $\alpha$-NDCG & ERR-IA & $\alpha$-NDCG \\
    \midrule
	Uniform & 0.2219 & 0.2835 & 0.4561 & 0.5793 \\
	Source-type & 0.2381 & 0.2905 & \textbf{0.4570} & \textbf{0.5832}  \\
	Individual & \textbf{0.2518}{$^\dag$} & \textbf{0.3064}{$^\dag$} & 0.4562 & \textbf{0.5832} \\
    \bottomrule
  \end{tabular}
  \label{table:src_imp}
  \vspace*{-0.3\baselineskip}
\end{table}

%% file: sigir2017-tasks-06.tex
\section{Conclusions}
\label{sec:concl}

In this paper, we have addressed the task of generating query suggestions that can assist users in completing their tasks. 
We have proposed a probabilistic generative framework with four components: source importance, document importance, keyphrase relevance, and query suggestions.  We have proposed and experimentally compared various alternatives for these components.

One important element, missing from our current model, is the representation of specific subtasks.  As a next step, we plan to cluster query suggestions together that belong to the same subtask. This would naturally enable us to provide diversified query suggestions.   

%% file: 00paper.bbl
%%% -*-BibTeX-*-
%%% Do NOT edit. File created by BibTeX with style
%%% ACM-Reference-Format-Journals [18-Jan-2012].

\begin{thebibliography}{00}

%%% ====================================================================
%%% NOTE TO THE USER: you can override these defaults by providing
%%% customized versions of any of these macros before the \bibliography
%%% command.  Each of them MUST provide its own final punctuation,
%%% except for \shownote{}, \showDOI{}, and \showURL{}.  The latter two
%%% do not use final punctuation, in order to avoid confusing it with
%%% the Web address.
%%%
%%% To suppress output of a particular field, define its macro to expand
%%% to an empty string, or better, \unskip, like this:
%%%
%%% \newcommand{\showDOI}[1]{\unskip}   % LaTeX syntax
%%%
%%% \def \showDOI #1{\unskip}           % plain TeX syntax
%%%
%%% ====================================================================

\ifx \showCODEN    \undefined \def \showCODEN     #1{\unskip}     \fi
\ifx \showDOI      \undefined \def \showDOI       #1{{\tt DOI:}\penalty0{#1}\ }
  \fi
\ifx \showISBNx    \undefined \def \showISBNx     #1{\unskip}     \fi
\ifx \showISBNxiii \undefined \def \showISBNxiii  #1{\unskip}     \fi
\ifx \showISSN     \undefined \def \showISSN      #1{\unskip}     \fi
\ifx \showLCCN     \undefined \def \showLCCN      #1{\unskip}     \fi
\ifx \shownote     \undefined \def \shownote      #1{#1}          \fi
\ifx \showarticletitle \undefined \def \showarticletitle #1{#1}   \fi
\ifx \showURL      \undefined \def \showURL       #1{#1}          \fi
% The following commands are used for tagged output and should be
% invisible to TeX
\providecommand\bibfield[2]{#2}
\providecommand\bibinfo[2]{#2}
\providecommand\natexlab[1]{#1}
\providecommand\showeprint[2][]{arXiv:#2}

\bibitem[\protect\citeauthoryear{Andolina, Klouche, Peltonen, Hoque, Ruotsalo,
  Cabral, Klami, Glowacka, Flor{\'e}en, and Jacucci}{Andolina
  et~al\mbox{.}}{2015}]%
        {Andolina:2015:ISP}
\bibfield{author}{\bibinfo{person}{Salvatore Andolina}, \bibinfo{person}{Khalil
  Klouche}, \bibinfo{person}{Jaakko Peltonen}, \bibinfo{person}{Mohammad~E.
  Hoque}, \bibinfo{person}{Tuukka Ruotsalo}, \bibinfo{person}{Diogo Cabral},
  \bibinfo{person}{Arto Klami}, \bibinfo{person}{Dorota Glowacka},
  \bibinfo{person}{Patrik Flor{\'e}en}, {and} \bibinfo{person}{Giulio
  Jacucci}.} \bibinfo{year}{2015}\natexlab{}.
\newblock \showarticletitle{IntentStreams: Smart Parallel Search Streams for
  Branching Exploratory Search}. In \bibinfo{booktitle}{{\em Proc. of IUI}}.
  \bibinfo{pages}{300--305}.
\newblock


\bibitem[\protect\citeauthoryear{Awadallah, White, Pantel, Dumais, and
  Wang}{Awadallah et~al\mbox{.}}{2014}]%
        {HassanAwadallah:2014:SCS}
\bibfield{author}{\bibinfo{person}{Ahmed~H. Awadallah},
  \bibinfo{person}{Ryen~W. White}, \bibinfo{person}{Patrick Pantel},
  \bibinfo{person}{Susan~T. Dumais}, {and} \bibinfo{person}{Yi-Min Wang}.}
  \bibinfo{year}{2014}\natexlab{}.
\newblock \showarticletitle{Supporting Complex Search Tasks}. In
  \bibinfo{booktitle}{{\em Proc. of CIKM}}. \bibinfo{pages}{829--838}.
\newblock


\bibitem[\protect\citeauthoryear{Balog}{Balog}{2015}]%
        {Balog:2015:TCE}
\bibfield{author}{\bibinfo{person}{Krisztian Balog}.}
  \bibinfo{year}{2015}\natexlab{}.
\newblock \showarticletitle{Task-completion Engines: A Vision with a Plan}. In
  \bibinfo{booktitle}{{\em Proc. of the 1st International Workshop on
  Supporting Complex Search Tasks}}.
\newblock


\bibitem[\protect\citeauthoryear{Bhatia, Majumdar, and Mitra}{Bhatia
  et~al\mbox{.}}{2011}]%
        {Bhatia:2011:QSA}
\bibfield{author}{\bibinfo{person}{Sumit Bhatia}, \bibinfo{person}{Debapriyo
  Majumdar}, {and} \bibinfo{person}{Prasenjit Mitra}.}
  \bibinfo{year}{2011}\natexlab{}.
\newblock \showarticletitle{Query Suggestions in the Absence of Query Logs}. In
  \bibinfo{booktitle}{{\em Proc. of SIGIR}}. \bibinfo{pages}{795--804}.
\newblock


\bibitem[\protect\citeauthoryear{Boldi, Bonchi, Castillo, Donato, Gionis, and
  Vigna}{Boldi et~al\mbox{.}}{2008}]%
        {Boldi:2008:QGM}
\bibfield{author}{\bibinfo{person}{Paolo Boldi}, \bibinfo{person}{Francesco
  Bonchi}, \bibinfo{person}{Carlos Castillo}, \bibinfo{person}{Debora Donato},
  \bibinfo{person}{Aristides Gionis}, {and} \bibinfo{person}{Sebastiano
  Vigna}.} \bibinfo{year}{2008}\natexlab{}.
\newblock \showarticletitle{The Query-flow Graph: Model and Applications}. In
  \bibinfo{booktitle}{{\em Proc. of CIKM}}. \bibinfo{pages}{609--618}.
\newblock


\bibitem[\protect\citeauthoryear{Craswell and Szummer}{Craswell and
  Szummer}{2007}]%
        {Craswell:2007:RWC}
\bibfield{author}{\bibinfo{person}{Nick Craswell} {and} \bibinfo{person}{Martin
  Szummer}.} \bibinfo{year}{2007}\natexlab{}.
\newblock \showarticletitle{Random walks on the click graph}. In
  \bibinfo{booktitle}{{\em Proc. of SIGIR}}. \bibinfo{pages}{239--246}.
\newblock


\bibitem[\protect\citeauthoryear{Garigliotti and Balog}{Garigliotti and
  Balog}{2016}]%
        {Garigliotti:2016:UST}
\bibfield{author}{\bibinfo{person}{Dar{\'{i}}o Garigliotti} {and}
  \bibinfo{person}{Krisztian Balog}.} \bibinfo{year}{2016}\natexlab{}.
\newblock \showarticletitle{{The University of Stavanger at the TREC 2016 Tasks
  Track}}. In \bibinfo{booktitle}{{\em Proc. of TREC}}.
\newblock


\bibitem[\protect\citeauthoryear{Hagen, G{\"{o}}ring, Keil, Anifowose, Othman,
  and Stein}{Hagen et~al\mbox{.}}{2015}]%
        {Hagen:2015:WAT}
\bibfield{author}{\bibinfo{person}{Matthias Hagen}, \bibinfo{person}{Steve
  G{\"{o}}ring}, \bibinfo{person}{Magdalena Keil}, \bibinfo{person}{Olaoluwa
  Anifowose}, \bibinfo{person}{Amir Othman}, {and} \bibinfo{person}{Benno
  Stein}.} \bibinfo{year}{2015}\natexlab{}.
\newblock \showarticletitle{Webis at {TREC} 2015: Tasks and Total Recall
  Tracks}. In \bibinfo{booktitle}{{\em Proc. of {TREC}}}.
\newblock


\bibitem[\protect\citeauthoryear{Hagen, Kiesel, Adineh, Alahyari, Fatehifar,
  Bahrami, Fichtl, and Stein}{Hagen et~al\mbox{.}}{2016}]%
        {Hagen:2016:WAT}
\bibfield{author}{\bibinfo{person}{Matthias Hagen}, \bibinfo{person}{Johannes
  Kiesel}, \bibinfo{person}{Payam Adineh}, \bibinfo{person}{Masoud Alahyari},
  \bibinfo{person}{Ehsan Fatehifar}, \bibinfo{person}{Arafeh Bahrami},
  \bibinfo{person}{Pia Fichtl}, {and} \bibinfo{person}{Benno Stein}.}
  \bibinfo{year}{2016}\natexlab{}.
\newblock \showarticletitle{{Webis at TREC 2016: Tasks, Total Recall, and Open
  Search Tracks}}. In \bibinfo{booktitle}{{\em Proc. of TREC}}.
\newblock


\bibitem[\protect\citeauthoryear{Jiang, He, Han, Yue, and Ni}{Jiang
  et~al\mbox{.}}{2012}]%
        {Jiang:2012:CEQ}
\bibfield{author}{\bibinfo{person}{Jiepu Jiang}, \bibinfo{person}{Daqing He},
  \bibinfo{person}{Shuguang Han}, \bibinfo{person}{Zhen Yue}, {and}
  \bibinfo{person}{Chaoqun Ni}.} \bibinfo{year}{2012}\natexlab{}.
\newblock \showarticletitle{Contextual Evaluation of Query Reformulations in a
  Search Session by User Simulation}. In \bibinfo{booktitle}{{\em Proc. of
  CIKM}}. \bibinfo{pages}{2635--2638}.
\newblock


\bibitem[\protect\citeauthoryear{Kelly, Arguello, and Capra}{Kelly
  et~al\mbox{.}}{2013}]%
        {Kelly:2013:NWT}
\bibfield{author}{\bibinfo{person}{Diane Kelly}, \bibinfo{person}{Jaime
  Arguello}, {and} \bibinfo{person}{Robert Capra}.}
  \bibinfo{year}{2013}\natexlab{}.
\newblock \showarticletitle{{NSF} Workshop on Task-based Information Search
  Systems}.
\newblock \bibinfo{journal}{{\em SIGIR Forum\/}} \bibinfo{volume}{47},
  \bibinfo{number}{2} (\bibinfo{year}{2013}), \bibinfo{pages}{116--127}.
\newblock


\bibitem[\protect\citeauthoryear{Kelly, Gyllstrom, and Bailey}{Kelly
  et~al\mbox{.}}{2009}]%
        {Kelly:2009:CQT}
\bibfield{author}{\bibinfo{person}{Diane Kelly}, \bibinfo{person}{Karl
  Gyllstrom}, {and} \bibinfo{person}{Earl~W. Bailey}.}
  \bibinfo{year}{2009}\natexlab{}.
\newblock \showarticletitle{A Comparison of Query and Term Suggestion Features
  for Interactive Searching}. In \bibinfo{booktitle}{{\em Proc. of SIGIR}}.
  \bibinfo{pages}{371--378}.
\newblock


\bibitem[\protect\citeauthoryear{Kruschwitz, Lungley, Albakour, and
  Song}{Kruschwitz et~al\mbox{.}}{2013}]%
        {Kruschwitz:2013:DQS}
\bibfield{author}{\bibinfo{person}{Udo Kruschwitz}, \bibinfo{person}{Deirdre
  Lungley}, \bibinfo{person}{M-Dyaa Albakour}, {and} \bibinfo{person}{Dawei
  Song}.} \bibinfo{year}{2013}\natexlab{}.
\newblock \showarticletitle{Deriving query suggestions for site search}.
\newblock \bibinfo{journal}{{\em {JASIST}\/}} \bibinfo{volume}{64},
  \bibinfo{number}{10} (\bibinfo{year}{2013}), \bibinfo{pages}{1975--1994}.
\newblock


\bibitem[\protect\citeauthoryear{Lucchese, Orlando, Perego, Silvestri, and
  Tolomei}{Lucchese et~al\mbox{.}}{2013}]%
        {Lucchese:2013:DTS}
\bibfield{author}{\bibinfo{person}{Claudio Lucchese},
  \bibinfo{person}{Salvatore Orlando}, \bibinfo{person}{Raffaele Perego},
  \bibinfo{person}{Fabrizio Silvestri}, {and} \bibinfo{person}{Gabriele
  Tolomei}.} \bibinfo{year}{2013}\natexlab{}.
\newblock \showarticletitle{Discovering Tasks from Search Engine Query Logs}.
\newblock \bibinfo{journal}{{\em ACM Trans. Inf. Syst.\/}}
  \bibinfo{volume}{31}, \bibinfo{number}{3}, Article \bibinfo{articleno}{14}
  (\bibinfo{year}{2013}), \bibinfo{numpages}{43}~pages.
\newblock


\bibitem[\protect\citeauthoryear{Medelyan}{Medelyan}{2015}]%
        {Medelyan:2015:RMT}
\bibfield{author}{\bibinfo{person}{Alyona Medelyan}.}
  \bibinfo{year}{2015}\natexlab{}.
\newblock \bibinfo{title}{Modified {RAKE} algorithm}.
\newblock
  \bibinfo{howpublished}{\url{https://github.com/zelandiya/RAKE-tutorial}}.
  (\bibinfo{year}{2015}).
\newblock
\newblock
\shownote{Accessed: 2017-01-23.}


\bibitem[\protect\citeauthoryear{Ozertem, Chapelle, Donmez, and
  Velipasaoglu}{Ozertem et~al\mbox{.}}{2012}]%
        {Ozertem:2012:LSM}
\bibfield{author}{\bibinfo{person}{Umut Ozertem}, \bibinfo{person}{Olivier
  Chapelle}, \bibinfo{person}{Pinar Donmez}, {and} \bibinfo{person}{Emre
  Velipasaoglu}.} \bibinfo{year}{2012}\natexlab{}.
\newblock \showarticletitle{Learning to Suggest: A Machine Learning Framework
  for Ranking Query Suggestions}. In \bibinfo{booktitle}{{\em Proc. of SIGIR}}.
  \bibinfo{pages}{25--34}.
\newblock


\bibitem[\protect\citeauthoryear{Tran, Schwarz, Nieder{\'e}e, Maus, and
  Kanhabua}{Tran et~al\mbox{.}}{2016}]%
        {Tran:2016:FNM}
\bibfield{author}{\bibinfo{person}{Tuan~A. Tran}, \bibinfo{person}{Sven
  Schwarz}, \bibinfo{person}{Claudia Nieder{\'e}e}, \bibinfo{person}{Heiko
  Maus}, {and} \bibinfo{person}{Nattiya Kanhabua}.}
  \bibinfo{year}{2016}\natexlab{}.
\newblock \showarticletitle{The Forgotten Needle in My Collections: Task-Aware
  Ranking of Documents in Semantic Information Space}. In
  \bibinfo{booktitle}{{\em Proc. of CHIIR}}. \bibinfo{pages}{13--22}.
\newblock


\bibitem[\protect\citeauthoryear{Verma, Kanoulas, Yilmaz, Mehrotra, Carterette,
  Craswell, and Bailey}{Verma et~al\mbox{.}}{2016}]%
        {Verma:2016:OTT}
\bibfield{author}{\bibinfo{person}{Manisha Verma}, \bibinfo{person}{Evangelos
  Kanoulas}, \bibinfo{person}{Emine Yilmaz}, \bibinfo{person}{Rishabh
  Mehrotra}, \bibinfo{person}{Ben Carterette}, \bibinfo{person}{Nick Craswell},
  {and} \bibinfo{person}{Peter Bailey}.} \bibinfo{year}{2016}\natexlab{}.
\newblock \showarticletitle{Overview of the {TREC} Tasks Track 2016}. In
  \bibinfo{booktitle}{{\em Proc. of {TREC}}}.
\newblock


\bibitem[\protect\citeauthoryear{White and Huang}{White and Huang}{2010}]%
        {White:2010:ASR}
\bibfield{author}{\bibinfo{person}{Ryen~W. White} {and} \bibinfo{person}{Jeff
  Huang}.} \bibinfo{year}{2010}\natexlab{}.
\newblock \showarticletitle{Assessing the Scenic Route: Measuring the Value of
  Search Trails in Web Logs}. In \bibinfo{booktitle}{{\em Proc. of SIGIR}}.
  \bibinfo{pages}{587--594}.
\newblock


\bibitem[\protect\citeauthoryear{Yilmaz, Verma, Mehrotra, Kanoulas, Carterette,
  and Craswell}{Yilmaz et~al\mbox{.}}{2015}]%
        {Yilmaz:2015:OTT}
\bibfield{author}{\bibinfo{person}{Emine Yilmaz}, \bibinfo{person}{Manisha
  Verma}, \bibinfo{person}{Rishabh Mehrotra}, \bibinfo{person}{Evangelos
  Kanoulas}, \bibinfo{person}{Ben Carterette}, {and} \bibinfo{person}{Nick
  Craswell}.} \bibinfo{year}{2015}\natexlab{}.
\newblock \showarticletitle{Overview of the {TREC} 2015 Tasks Track}. In
  \bibinfo{booktitle}{{\em Proc. of {TREC}}}.
\newblock


\end{thebibliography}
